# Revisiting the Effect of *f*-Functions in Predicting the Right Reaction Mechanism for Hypervalent Iodine Reagents

Tian-Yu Sun,[a] Kai Chen,[b,c]* Huakang Zhou,[d] Tingting You,[a]* Penggang Yin,[a] and Xiao Wang[e]*

a.  Key Laboratory of Bio-Inspired Smart Interfacial Science and Technology of Ministry of Education, School of Chemistry, Beihang University, Beijing 100191, P. R. China. E-mail: youtt@buaa.edu.cn.
b.  School of Chemistry and Chemical Engineering, Central South University, Changsha 410083, P. R. China.
c.  Lab of Computational Chemistry and Drug Design, State Key Laboratory of Chemical Oncogenomics, Peking University Shenzhen Graduate School, Shenzhen, 518055, P. R. China. E-mail: chenkaink@gmail.com.
d.  School of Materials Science and Engineering, Central South University of Forestry and Technology, Changsha 410004, P. R. China.
e.  Center for Computational Quantum Physics, Flatiron Institute, New York, NY 10010, USA. E-mail: xwang@flatironinstitute.org.

## Abstract

To understand the effect of *f*-functions in predicting the right reaction mechanism for hypervalent iodine reagents, we adopt the Ahlrichs basis set family def2-SVP and def2-TZVP to revisit the potential energy surfaces of **IBX**-mediated oxidation and **Togni I**'s isomerisation. Our results further prove that *f*-functions (in either Pople, Dunning, or Ahlrichs basis set series) are indispensable to predict the correct rate-determining step of hypervalent iodine reagents. The *f*-functions have a significant impact on the predicted reaction barriers for processes involving the I-X (X = O, OH, $CF_3$, etc.) bond cleavage and formation, e.g. in the reductive elimination step or the hypervalent twist step. We furthermore explore two hypervalent twist modes that account for the different influences of *f*-functions for **IBX** and **Togni I**. Our findings may be helpful for theoretical chemists to appropriately study the reaction mechanism of hypervalent iodine reagents.

# 1. Introduction

Hypervalent iodine reagents,[1-3] such as 2-iodoxybenzoic acid (**IBX**) and Togni's reagent I (**Togni I**) (see **Scheme 1**), have attracted broad interests in recent years because of their economic feasibility, eco-friendliness, and low toxicity, and both theoretical[4-9] and experimental[10-13] studies of these reagents has undergone an explosive growth.

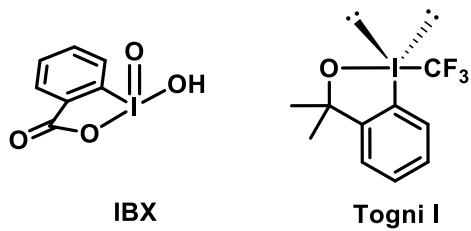

**Scheme 1**. The structures of **IBX** and **Togni I**.

**IBX** is an important oxidant due to its ideal chemo-selectivity and mild reaction conditions.[14] The **IBX**-mediated oxidation of alcohol mainly includes two steps, i.e. (1) hypervalent twist where ligands attached to iodine are rearranged in a coordinated motion and (2) reductive elimination that involves the C−H bond cleavage. Identifying its rate-determining step (RDS) has both scientific and engineering importance. Earlier density functional theory (DFT) study[15] determined the hypervalent twist as the RDS using the MPW1K functional and the LACV3P** basis set,[16] which is a combination of Pople's 6-311G** basis set[14] and the LANL2DZ effective core.[17] Same conclusion has been drawn later by other researchers using various functionals with basis sets similar to LACV3P**.[18-21] However, this conclusion is inconsistent with the Corey's kinetic isotope effect (KIE) experiments.[22]

In 2017, our theoretical study employing a wide range of basis sets at the level of both DFT and density cumulant theory (DCT) revealed that the RDS of **IBX**-mediated oxidation is strongly sensitive to the choice of basis sets.[23] With LACV3P** as well as the Dunning basis sets without *f*-functions (such as cc-pVDZ(-PP) and aug-cc-pVDZ(-PP), where the (-PP) suffix represents pseudopotential for iodine), the RDS was predicted to be the hypervalent twist step (**TS1**). With more complete Dunning basis sets, such as cc-pVTZ(-PP), cc-PVQZ(-PP), aug-cc-pVTZ(-PP), and aug-cc-PVQZ(-PP),[24-29] all tested functionals including MPM1K,[30] M06-2X,[31] ωB97X-D[32] and B3LYP[33] determined the reductive elimination (**TS2**) as the RDS; this conclusion remains unchanged as we approach the basis set limit.[23] We have attributed the two contradictory results to the effect of *f*-functions in the basis sets.

Another important hypervalent iodine reagent, **Togni I**, has been successfully applied as an oxidant to synthesize trifluoromethylated compounds in pharmaceutical and agrochemical industries.[34-35] Our previous work[36] has showed that **Togni I** can also undergo a two-step mechanism, i.e. the hypervalent twist (**TS3**) and the reductive elimination, to form its lower-energy ether isomer. Similar to the **IBX**-mediated oxidation reaction, **Togni I**'s isomerisation was determined to feature an RDS of reductive elimination (**TS4**) at the B3LYP/aug-cc-pVTZ(-PP) level of theory. However, whether the *f*-functions play an essential role in the determination of the RDS remains to be confirmed.

In order to generalize our findings on the effect of *f*-functions, in the present research we adopt a different basis set family, i.e. def2-SVP and def2-TZVP,[37-39] to revisit the reactions of **IBX**-mediated oxidation and **Togni I**'s isomerisation. The layout of the article is as follows. In Section 2, we explain the basis sets and methods used in our work. In Section 3, we discuss the effect of *f*-functions on reaction mechanism of the two hypervalent iodine reagents, and further analyse the origins of such effect. In Section 4, we summarize our results.

## 2. Computational Details

Compared to the cc-pVDZ(-PP) and cc-pVTZ(-PP) basis sets, def2-SVP and def2-TZVP have respectively similar components of contracted orbitals with a smaller number of primitive functions. The def2-SVP basis set can be described as (4s1p/2s1p) for H, (7s4p1d/3s2p1d) for C and O, and (10s7p6d/4s4p2d) for I, while the def2-TZVP basis set can be described as (5s1p/3s1p) for H, (11s6p2d1f/5s3p2d1f) for C and O, and (11s10p8d2f/6s5p3d2f) for I. Note that *f*-functions for C, O, and I are provided in def2-TZVP but not contained in def2-SVP. In the present work, M06-2X was chosen as the DFT functional for all computations because it has been recommended for the study of main-group thermochemistry and kinetics.[31]

## 3. Results and Discussion

### 3.1 Effect of *f*-functions for IBX

The potential energy surface (PES) of the oxidation reaction for **IBX** is shown in **Figure 1**. Two curves (blue and green) are obtained at the M06-2X/def2-SVP and M06-2X/def2-TZVP level of theory,[40] respectively. Additional curve (red) from M06-2X with the aug-cc-pVTZ(-PP) basis set is provided as a calibration. With the def2-SVP basis set, the hypervalent twist (**TS1**) is incorrectly predicted as RDS (**Figure 1**), with the **TS1** energy barrier 3.5(=19.1-15.6) kcal/mol higher than that of reductive elimination (**TS2**). On the contrary, when the def2-TZVP basis set is adopted, the RDS is correctly predicted, with **TS2** energy higher than **TS1** by 8.0 kcal/mol. **Figure 1** also shows that the def2-TZVP energy barriers qualitatively reproduced the aug-cc-pVTZ(-PP) results (differ by 0.1 kcal/mol for **TS1** and 2.2 kcal/mol for **TS2**). Combining with our previous work,[23] we have investigated three different basis set families, including Pople's 6-311G**, Dunning's correlation consistent basis sets, and Ahlrichs' improved "def-bases", and confirmed that the *f*-functions are critical to predict the correct RDS for the reactions of **IBX**.

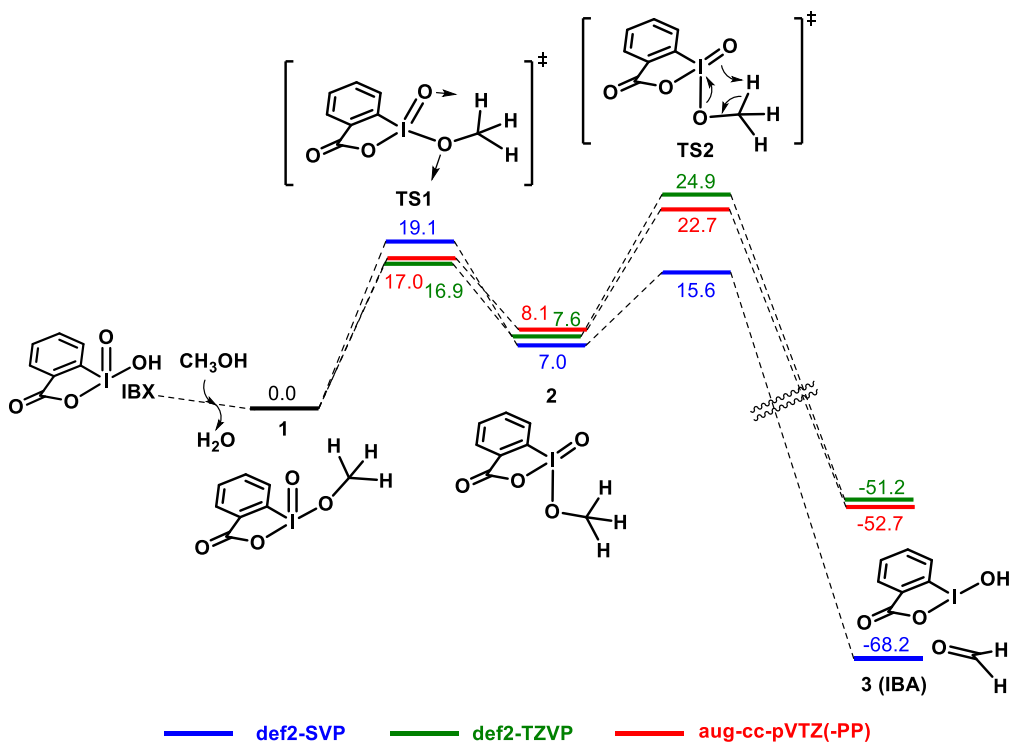

**Figure 1.** The PESs of **IBX**-mediated oxidation of methanol by def2-SVP, def2-TZVP, aug-cc-pVTZ(-PP). The free energy values are in kcal/mol.

### 3.2 Effect of *f*-functions for Togni I

The isomerisation reaction path of **Togni I** is shown in **Figure 2.** With the smaller def2-SVP basis set (without *f*-functions, blue curve), the hypervalent twist (**TS3**) is predicted as the RDS, with an energy barrier 11 kcal/mol higher than that of reductive elimination (**TS4**). In the contrast, def2-TZVP (with *f*-functions, green curve) predicts the reductive elimination (**TS4**) as the RDS, whose energy barrier is 3 kcal/mol lower than that of **TS3**. Again, the PES predicted by def2-TZVP almost reproduces the aug-cc-pVTZ(-PP) results (red curve) within an energy difference of 2 kcal/mol. These results also indicate that the *f*-functions are critical to predict the correct RDS for the reactions of **Togni I**.

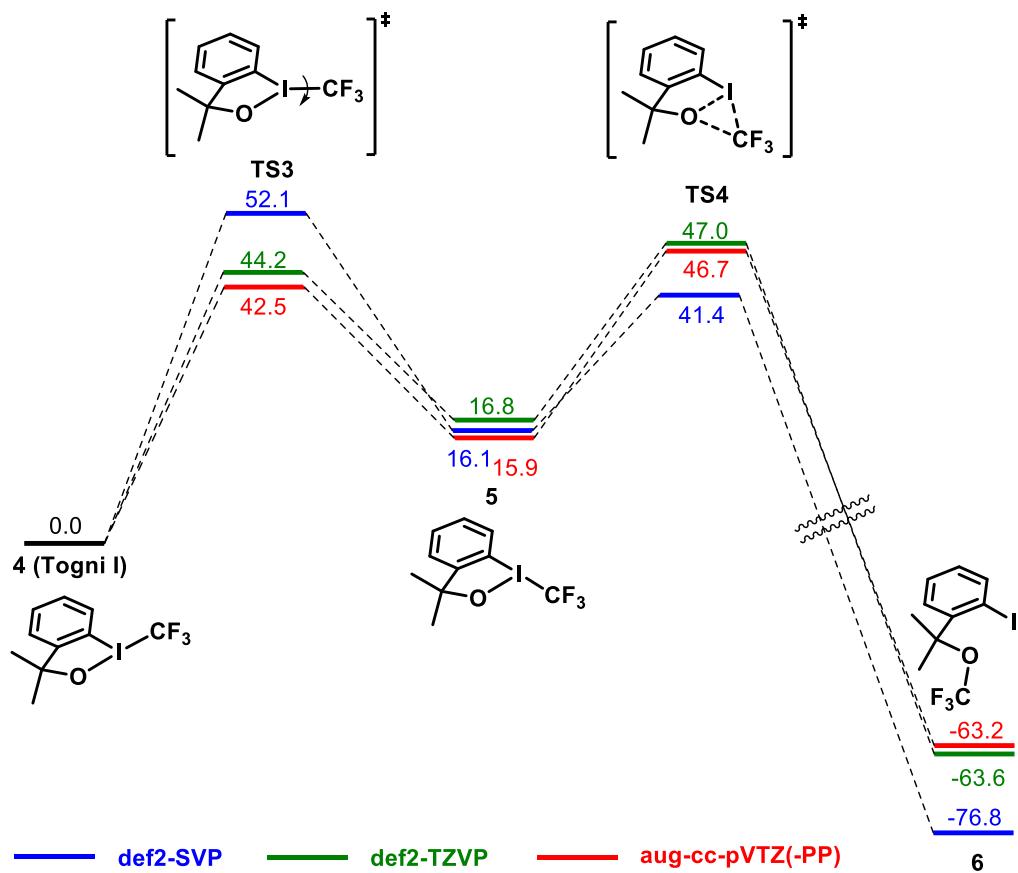

**Figure 2.** The isomerization PESs of **Togni I** by def2-SVP, def2-TZVP, aug-cc-pVTZ(-PP). The free energy values are in kcal/mol.

### 3.3 Origins of different *f*-function dependence

The wrong RDS predicted by def2-SVP in **Figures 1** and **2** can simply be attributed to both the underestimation of the energy barrier for the reductive elimination step (**TS2** or **TS4**) and the overestimation of the energy barrier for the hypervalent twist step (**TS1** or **TS3**). Next, we individually analyse the two steps due to their contrasting dependence on the basis sets.

As one source of error, the underestimation of **TS2** by 7 kcal/mol relative to the aug-cc-pVTZ(-PP) calibration is the main reason for the failure of the prediction of RDS for **IBX** (see **Figure 1**). This is directly related the more severe underestimation (by 16 kcal/mol) of the energy for the elimination product **3**. In other words, the def2-SVP basis sets cannot even qualitatively predict the correct thermodynamics. As halfway through the bond-breaking process, the transition state **TS2** is underestimated as well. Interestingly, the energy deviation in **TS2** is about half as much as in **3**. Likewise, in **Figure 2**, def2-SVP underestimates the reaction energy of product **6** by 14 kcal/mol compared to that predicted by aug-cc-pVTZ(-PP), and causes the related reaction barrier of **TS4** to deviate from the calibration by 5 kcal/mol. Thus, the *f*-functions are shown to be critical to predict the correct energies for the process of the I–O (**Figure 1**) or I–C (**Figure 2**) bond breaking. In a 2011 paper by Truhlar et al., the def2-SVP basis set was used along with various DFT methods to determine the acid dissociation energy for $H_3AsO_4$, and the mean unsigned errors (MUE) was found to be 10.5 kcal/mol. However, when the def2-TZVP basis set was used, the MUE dropped off to 2.0 kcal/mol.[41] This may imply that *f*-basis functions play important roles for these reactions involving fourth- and fifth-period main group elements.

Another source of error that leads to incorrect identification of RDS is the overestimation of energy barrier of the hypervalent twist step (**TS1** in **Figure 1**, **TS3** in **Figure 2**). In contrast to their significant effects on both **TS2** and **TS4**, the *f*-basis functions are only important for the hypervalent twist step in **Togni I** reaction (**TS3**, **Figure 2**), but have less influence on that in **IBX** reaction (**TS1**, **Figure 1**). This difference may be explained by the different hypervalent twist pathways for the two reactions, which result in different geometries of the transition states. Our computations show that the hypervalent twist step is realized via an *equatorial* position in **IBX** but via an *apical* position in **Togni I** (**Scheme 2**). The equivalent terms *out-of-plane* and *in-plane* were used, respectively, in Lüthi's work.[42]

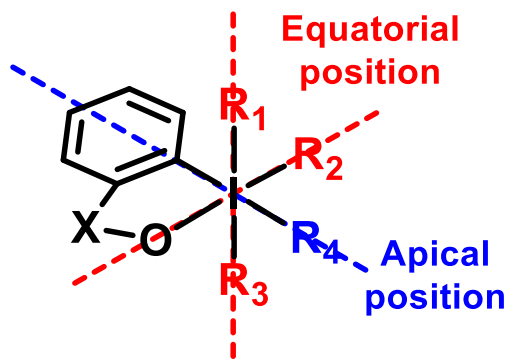

**Scheme 2.** Apical and equatorial positions in hypervalent iodine reagents.

As shown in **Scheme 3**, for the **IBX** reaction the hypervalent twist from **1** to the intermediate **2** is via an internal rotation around the I–Ph bond axis. In **TS1**, the −OCH$_3$ and =O groups are in the equatorial positions (**Scheme 3a**), leaving the I–OCH$_3$ and I=O distances almost unchanged. The endocyclic I-O bond is highly ionic, as indicated by the NBO analysis,[43] and thus, in this twist mode there is no bond breaking involved, which is believed to have weak dependence on the basis sets.[44] As a result, the twist energy barrier of **TS1** is relatively low (< 20 kcal/mol) and not significantly influenced by the *f*-functions.

However, for the **Togni I** reaction, in the hypervalent twist transition state **TS3**, the CF$_3$ group is predicted to move towards the apical position (**Scheme 3b**). Since the CF$_3$ group tending to the apical position must have strong interaction with the two lone pairs on the I atom (see **Scheme 1**), the twist step in this reaction requires higher energy (> 40 kcal/mol), and the I–CF$_3$ distance in **TS3** gets lengthened by about 30%. Because the basis set overlap between I and CF$_3$ depends on the I–CF$_3$ distance, it is not surprising that the effect of *f*-basis functions is more noticeable for **TS3** than for **TS1**.

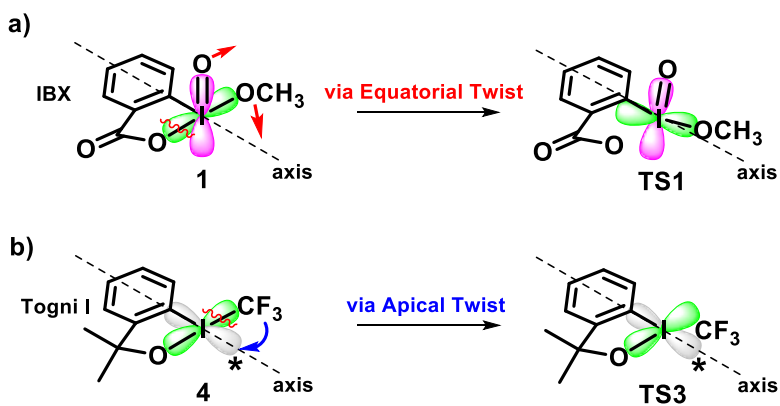

**Scheme 3**. The NBO analysis for the **TS1** of **IBX** and the **TS3** of **Togni I**. The orbitals with asterisks in b) (in the grey color) represent anti-bonding orbitals.

## 4. Conclusions

In summary, we adopted the Ahlrichs def2-SVP and def2-TZVP basis sets to study the two-step reactions of **IBX** and **Togni I**. Similar to our previous work with the Pople and Dunning basis set families, our present results confirm that the *f*-functions in Ahlrichs basis sets are important to predict the right rate-determining step (RDS).

One reason for the failure of the smaller basis sets (without *f*-functions) in identifying RDS is that they tend to predict significantly lower energy barriers for the I-C and I-O bond breaking (**TS2** and **TS4**) in the reductive elimination step, compared to the results from a larger basis set (with *f*-functions). And such underestimation is associated with their incapability to correctly predict reaction energies.

Another reason for the prediction of wrong RDS involves the hypervalent twist step. The reaction barrier of **TS1** (< 20 kcal/mol) is much lower in energy than that of **TS3** (> 40 kcal/mol), and the effect of *f*-functions is different on the hypervalent twist between **IBX** and **Togni I**. This could be related to the different hypervalent twist pathways (equatorial twist pathway and apical twist pathway) and bond length changes.

Since theoretical studies for the reactions of hypervalent iodine reagents provide significant guidance for experiments, we hope our findings on the effect of *f*-functions could facilitate the determination of correct reaction mechanism for more hypervalent iodine reagents.

## Conflicts of interest
There are no conflicts to declare.

## Acknowledgements

This work is supported by NSFC (31890774, 31890770), Guangdong NSF (2016A030310433), Hunan NSF (2018JJ3868), the Youth Scientific Research Foundation of Central South University of Forestry and Technology (QJ2017005B), and Peking University Shenzhen Graduate School. TYS acknowledges Li-Ying Pan for helpful discussions. KC acknowledges the National Supercomputer Center in Guangzhou for computer resources and Lü Liang for the technical support. The Flatiron Institute is a division of the Simons Foundation.